\documentclass[aps,prl,preprint,superscriptaddress]{revtex4-1}

\usepackage{graphicx}

\def\dif{{\mathrm d}}
\def\vecp{{\mathbf p}}
\def\vecr{{\mathbf r}}
\def\vecv{{\mathbf v}}
\def\vecx{{\mathbf x}}
\def\vF{{\mathbf F}}
\def\Ref#1{(\ref{#1})}
\def\rmin{a}
\def\ene{{\cal E}}

\begin{document}


\title{Classical molecular dynamics simulations of hydrogen plasmas and
       development of an analytical statistical model for computational validity
       assessment}


\author{M. A. Gigosos}
\email[]{gigosos@coyanza.opt.cie.uva.es}
\author{D. Gonz\'alez-Herrero}
\author{N. Lara}
\affiliation{Departamento de F\'isica Te\'orica, At\'omica y \'Optica,
	     Universidad de Valladolid, 47071 Valladolid, Spain}

\author{R. Florido}
\affiliation{iUNAT - Departamento de F\'isica,
	     Universidad de Las Palmas de Gran Canaria,
	     35017 Las Palmas de Gran Canaria,
	     Spain}

\author{A. Calisti}
\author{S. Ferri}
\author{B. Talin}
\affiliation{Aix Marseille Universit\'e, CNRS, PIIM, 13397 Marseille,
	     France}


\date{\today}

\begin{abstract}
    Classical molecular dynamics simulations of hydrogen plasmas
    have been performed with emphasis on the analysis of
    equilibration process. Theoretical basis of simulation model as
    well as numerically relevant aspects --such as the proper choice
    and definition of simulation units-- are discussed in detail,
    thus proving a
    thorough implementation of the computer simulation technique.
    Because of lack of experimental data, molecular dynamics
    simulations are often considered as
    \textit{idealized computational experiments}
    for benchmarking of theoretical models. However,
    these simulations are certainly challenging and consequently a
    validation procedure is also demanded. In this work we develop
    an analytical statistical equilibrium model for computational validity
    assessment of plasma particle dynamics simulations. Remarkable
    agreement between model and molecular dynamics results including
    a classical treatment of ionization-recombination mechanism is
    obtained for a wide range of plasma coupling parameter.
    Furthermore, the analytical model provides guidance to securely
    terminate simulation runs once the equilibrium stage has been
    reached, which in turn gives confidence on the statistics that
    potentially may be extracted from time-histories of simulated
    physical quantities.
\end{abstract}


\maketitle


\section{\label{intro}Introduction}
    Computer simulation is the discipline of designing an abstract model
    to reproduce the dynamics and behavior of an actual physical system,
    translating the model into a computer program, and analyzing the
    data obtained from the program execution. With the increase of
    computational power and data storage, computer simulations have
    proven to be a valuable tool in many different fields in physics
    because of its ability for solving complex problems just relying on
    fundamental first principles and barely using either physical or
    mathematical approximations. Results from computer simulations are
    often considered as \textit{idealized experiments}, where different
    effects can be artificially switch on and off to assess their
    potential impact, thus providing a deep insight of the underlying
    physics and a unique testbed for theory validation.

    In particular, the use of computer simulations to study the problem
    of broadening of spectral line shapes in plasmas has a long history,
    having significantly contributed to the development and improvement
    of theoretical models~\cite{stambulchik2010}. Nowadays, the theory
    of Stark broadening has matured enough and become one of the most
    important diagnostic tools for astrophysical and laboratory plasmas.
    However, some issues remain still open, e.g. line broadening theory
    has been validated using independent methods of extracting plasma
    conditions only for low-Z elements at particles densities below
    $10^{25}$~m$^{-3}$, disagreement between different approaches
    persist especially in describing the ion motion
    effects~\cite{stamm1979,stamm1986,calisti2014,ferri2014} and also
    discrepancies in the
    line shape calculations have been pointed out as the major source
    of uncertainty in the inferred plasma conditions from the analysis
    of K-shell spectra observed in opacity-related
    experiments~\cite{nagayama2016}. This scenario has stimulated the
    research on line broadening over the last few years and led to a
    series of dedicated workshops for detailed comparisons of
    computational and analytical methods in order to identify sources
    of discrepancies and set model validity
    ranges~\cite{stambulchik2013,stambulchik2014,rosato2017}. 
 
    Computer simulations applied to calculations of Stark-broadened
    line shapes follow a three-steps scheme~\cite{stambulchik2010}.
    The first one consists of simulating the plasma particle dynamics,
    i.e. the motion of electrons, ions and neutrals as result of their
    mutual interactions of electric nature. Particle dynamics simulations
    (PDS) provide information about the behavior and statistical
    properties of local electric microfields, which are ultimately
    responsible for the Stark broadening and shift of line transitions.
    In the second step, a representative statistical sample of time
    histories of the local electric microfield is used to numerically
    integrate the time-dependent Schr\"odinger's equation of the
    \textit{radiator}, i.e. the emitting ion or atom, and compute the
    dipole autocorrelation function. In the third step, the Fourier
    transform of the autocorrelation function, i.e. the power spectral
    density, is computed, which finally leads to the spectral line
    profile. In this process, the first step is the most challenging
    one, since the last two rely on PDS ability to provide a faithful
    picture of particle motion and an accurate representation of plasma
    equilibrium states, which is critical to determine the correct
    statistics of physical quantities. This work provides insight on the
    physical and numerical requirements needed for performing reliable PDS.

    Mainly, two different approaches have been used along time to simulate
    the plasma particle dynamics. The first one follows the independent
    particle approximation (IPA)~\cite{stamm1979,stamm1986,gigosos1987,
    hegerfeldt1988,cardenoso1989,gigosos1996,stambulchik2006,
    stambulchik2007b,gigosos2014}, i.e.
    particle interactions are neglected and they all move following
    straight-path trajectories. When computing time-histories of relevant
    physical quantities a Debye screened field is assumed to account for
    coupling effects. Obviously, IPA validity is limited to weakly-coupled
    plasmas. The second approach relies on molecular dynamics (MD)
    simulations. Now, interactions among all particles are explicitly
    included and calculations therefore are quite computationally
    demanding. On the upside, however, collective behaviors ---e.g. ion
    dynamic effects--- emerge in a natural way, the range of validity
    extends to strongly-coupled plasmas and, in the lack of experimental
    data, MD results are often considered as a reference to reveal model
    deficiencies and provide a valuable guidance for theory improvement.
    Classical MD simulations have been applied to the study of diverse
    statistical properties, particle correlation effects, and in particular
    to the investigation of plasma electric microfield
    distributions~\cite{hansen1978,hansen1981,hansen1983,fisher2001,
    nersisyan2005,sadykova2010,calisti2011,hauriege2015,hauriege2017}.
    Although mainly performed in the context of fully-ionized two-component
    plasmas, all this work enabled the study of electric microfield issues
    beyond the capability of most theoretical methods. Furthermore, full
    MD simulations have been used in several works to carry out elaborate
    calculations of Stark-broadened line profiles in hydrogenic
    plasmas~\cite{stambulchik2006,ferri2007,stambulchik2007,nersisyan2010}.
      
    Performing a simulation based on MD techniques is not a straightforward
    task. A thorough analysis and implementation of numerical and
    computational modeling of the physics involved are required to optimize
    the computational time and warrant reliable calculations. In particular,
    MD simulations have to deal with two pathologic problems:
    \textit{(i)} the simultaneous simulation of both light and heavy
    particles and
    \textit{(ii)} the requirement for the system to reach a stationary
    state in order to provide meaningful statistical samples of the
    relevant physical quantities. As discussed in Sec.~\ref{mdmodel},
    first issue demands a careful analysis of all characteristic time and
    length scales in the system and its consistent implementation into the
    numerical algorithms to solve the particle dynamics equations. Second
    issue is the most delicate one. At the beginning, when Coulomb
    interactions are switched on, the initial distribution of electrons and
    ions constitutes a plasma out of equilibrium and an exchange of kinetic
    and potential energies then takes place between particles throughout the
    simulation volume. Ideally, at the end of the relaxation phase,
    statistical measurements can provide an equilibrium temperature together
    with a density of ion-electron pairs ---i.e. recombined ions--- and
    populations of free electrons and ions which fully characterize the
    equilibrated plasma. In between, the plasma state is slowly evolving and
    quite undefined until it can be considered as stationary. Thus, in this
    work we develop a method to carefully control the approach to
    equilibrium allowing to know when this slow fluctuating evolution can be
    securely interrupted to get one of the expected set of particle
    positions and velocities, i.e. our technique provides a way to achieve
    a well defined plasma equilibrium state.

    Also, when simulating plasma particle dynamics, several
    papers~\cite{fisher2001,talin2001,talin2002,calisti2007,
    nersisyan2010,graziani2012} have discussed the difficulty to deal with
    the situation in which an electron
    is trapped by a charged ion ---which may restrict the model applicability
    to the weak-coupling regime---. In this context a few works used MD
    techniques to model the plasma ionization balance ---i.e. including the
    ionization-recombination mechanism---~\cite{calisti2009,calisti2011b,
    hauriege2013}.  In the simulation model described here
    ionization-recombination process is explicitly included within
    a classical framework, so that recombined ions
    and neutral pairs are actually native constituents of the final
    ionization balance and equilibrium state.
    Our model is therefore appropriate for the study of
    strongly-coupled plasmas beyond the fully-ionized scenario, which in turn
    makes it particularly useful for the calculation of Stark-broadened line
    shapes. Such study will be addressed on a forthcoming publication. Here,
    we first focus on demonstrating the robustness and internal consistency
    of the simulation technique. Thus, benchmarking of numerical algorithms
    and results are shown for hydrogen plasmas~\cite{natitesis}, although
    our technique can be indeed applied for modelling of general multicharged
    plasmas~\cite{diegotesis}.
    Within the framework of classical statistics, we developed an analytical
    model that mimics the idealized picture of a computer simulated hydrogen
    plasma and allows to obtain the corresponding equilibrium state for given
    conditions. In this regard, when compared with MD simulation results,
    the statistical model, firstly, provides a way to prove that a unique
    equilibrium state has been reached at the end of the relaxation phase
    and, secondly, leads to a practical definition of a classical atom,
    which in turn enables the proper definition of a criterion to classify
    the electron population in the plasma into trapped and free ones. With
    computer simulations considered as reference numerical experiments, our
    statistical equilibrium model represents a powerful tool to assess the
    computational validity of MD simulations and the accuracy of employed
    numerical methods. To the best of our knowledge, this is the first time
    in which such crossed-comparison is made. 

\section{\label{mdmodel}Molecular dynamics simulation model}

\begin{figure}[t]
    \includegraphics[width=0.6\columnwidth]{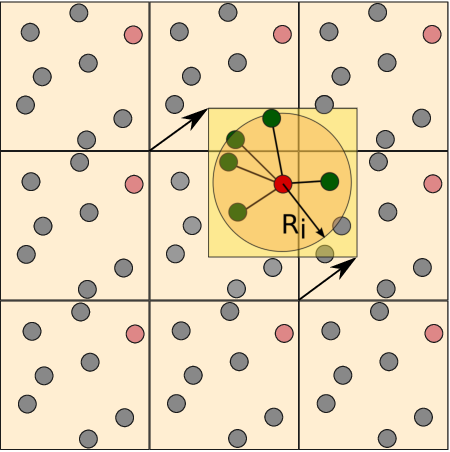}
    \caption{\label{FIG01}
    	2-D representation of the cubic simulation box with periodic
	boundary conditions. Each particle only interacts with others
	within a sphere of radius $R_I$.}
\end{figure}

    This work focuses on classical MD simulations of particle dynamics of
    hydrogen plasmas. In this framework, the simulation box is a cube of
    side $L$ containing $n_p$ electrons with mass $m_e$ and charge $-q$
    and $n_p$ ions with mass $m_i$ and charge $+q$. Boundary periodic
    conditions are assumed, i.e. when a particle leaves the box at a given
    velocity and direction, a particle of the same type enters from the
    opposite side with exactly the same velocity and direction. At this
    point it is convenient  to recall the definition of some global plasma
    parameters. Thus, for a hydrogen plasma characterized by a free
    electron density $N_e$ and an equilibrium temperature $T$, 
    $r_0 = \left(3/4\pi N_e\right)^{1/3}$ gives the average
    electron-electron distance, and the Debye length,
    $\lambda_D = \left(\varepsilon_0 kT/q^2N_e\right)^{1/2}$, measures
    the effective range of Coulomb interactions as a result of the plasma
    constituents coupling. Characteristic plasma time scale is given by
    $t_0  \equiv r_0/v_0$, where $v_0 \equiv \sqrt{2kT/m_e}$
    is the characteristic electron velocity. It is common to introduce the
    dimensionless coupling parameter
    $\rho=r_0/\lambda_D\propto N_e^{1/6}T^{-1/2}$, with $1/\rho^3$ giving
    the average number of free electrons within Debye's
    sphere~\cite{gigosos1996}. A frequent alternative definition of
    the coupling parameter is 
    $\Gamma = \frac{q^2}{4\pi\varepsilon_0 r_0}\frac{1}{2kT}$, which
    represents the ratio between typical Coulomb potential energy and
    particle kinetic energy. Both parameters satisfy $\Gamma=\rho^2/6$.
    We note that different combinations of $N_e$ and $T$ may lead to the
    same $\rho$ (or $\Gamma$) value. Some representative values are:
    a) for arc discharge plasmas, with $N_e\sim10^{22}$~m$^{-3}$ and
    $kT\sim1$~eV, $\rho\sim0.4$ ($\Gamma\sim3\times10^{-2}$),
    b) for representative tokamak conditions, $N_e\sim10^{18}$~m$^{-3}$ and
    $kT\sim1$~keV, $\rho\sim0.003$ ($\Gamma\sim1.5\times10^{-6}$), and
    c) for an inertial-fusion imploding plasma, with $N_e\sim10^{29}$~m$^{-3}$
    and $kT\sim1$~keV, $\rho\sim0.2$ ($\Gamma\sim7\times10^{-3}$). 
    In order to keep computational resources and cost within practical limits,
    each particle is assumed to interact only with charges within a sphere of
    radius $R_I=L/2$ centered at the particle --see Fig.~\ref{FIG01}--. This
    \emph{sphere of interaction} allows to remove anisotropic effects that
    naturally arise due to cubic shape of the whole enclosure. We notice that
    this assumption does not introduce any additional approximation in the
    simulation as long as $R_I\gg\lambda_D$. The latter condition actually
    sets a lower bound for the number of particles to be used in the
    simulation, i.e. $n_p\gg1/\rho^3$. In other words, plasma conditions,
    $N_e$ and $T$, determine a minimum number of particles to be included in
    the simulation. For instance, for $N_e\sim10^{29}$~m$^{-3}$ and
    $kT\sim1$~keV, $n_p\gg125$. 

\subsection{Regularized potential}\label{regpot}
    In order to avoid the collapse of a classical system of ions and electrons
    interacting through Coulomb forces, the attractive behavior of Coulomb
    potential should be modified at short distances, thus leading to a finite
    value at the origin. Such procedure is known as \emph{potential
    regularization} and has been extensively discussed in the literature.
    In summary, two alternatives have been proposed for the choice of a
    \emph{regularized} potential. The first one is the use of so-called quantum
    statistical potentials (QSPs)~\cite{kelbg1963,dunn1967,deutsch1977,
    norman1979,minoo1981,ebeling1999,filinov2003,ebeling2006}, which were devised
    to take into account short-range quantum effects and avoid divergences in
    statistical thermodynamics due to Coulomb potential singularity. QSPs were
    used for the first time in classical MD simulations to investigate hydrogen
    plasma properties in a strong-coupling regime~\cite{hansen1978,hansen1981,
    hansen1983}. The second alternative, see Refs.~\cite{calisti2007,calisti2009,
    calisti2011,calisti2011b}, has a phenomenological origin and was constructed
    to improve the modeling of ion population kinetics in MD simulations. While
    QSPs' behaviour at short distances typically depends on plasma temperature
    through the thermal de Broglie wavelength, the latter one is designed to
    match the corresponding ionization energy at the origin. Impact of using
    different type of potentials on statistical properties of dense hydrogen
    plasmas with impurities has been recently
    studied~\cite{hauriege2015,hauriege2017}. Neglecting the
    ionization-recombination mechanism, these works suggest that slow electric
    microfield distributions are rather insensitive to the potential alternatives
    and, therefore, such choice would have a small impact on the calculation of
    Stark-broadened line profiles. The reader interested on these topics is
    referred to given references for details. 
    
    Here we propose a phenomenological ion-electron potential with a quadratic
    behavior at short distances.  A similar model is employed to describe the
    nuclear interaction in the well-known relativistic self-consistent field
    Hartree-Fock {ATOM} package~\cite{amusia1997,chernysheva1999} and the more
    recent and widely-used Flexible Atomic Code (FAC)~\cite{gu2008} for
    spectroscopic-quality calculations of atomic structure. As shown in
    Sec.~\ref{eqmodel}, this choice has the major advantage of permitting to
    develop an analytical plasma equilibrium model, that will be further used to
    assess the computational validity of our simulation technique. Thus, the
    ion-electron potential energy $V_{ie}(r)$ is defined as 
    \begin{equation}\label{pot}
	V_{ie}(r) = \left\{\begin{array}{ll}
	    \displaystyle
            V_i\left[\frac{1}{3}\left(\frac{r}{a}\right)^2-1\right], &
	    r \le \rmin
            \\ \\
            \displaystyle
            -\frac{q^2}{4\pi\varepsilon_0}\frac{1}{r}, &
	    \rmin  < r \le R_I
            \\ \\
            \displaystyle
	    0, &  r > R_I
	\end{array}\right.
    \end{equation}
    with $V_i$ denoting the ionization energy, and 
    \begin{equation}\label{asize}
	\rmin = \frac{3}{2}\frac{q^2}{4\pi\varepsilon_0}\frac{1}{V_i} 
    \end{equation}
    being determined to satisfy continuity and derivability conditions.
    We note in turn that $\rmin$, which ultimately depends on the ionization
    energy $V_i$, provides an estimate for the characteristic atomic size.
    From a classical point of view, this potential corresponds to the case
    of having the ion charge uniformly distributed in the volume of a sphere
    with radius $a$, which is also permeable to point-like electrons. 

    The ion-electron force then results
    \begin{equation}\label{force}
	\vF_{ie} (r)= \left\{\begin{array}{ll}
	  \displaystyle
             -\frac{q^2}{4\pi\varepsilon_0 \rmin^3}\vecr, & r \le \rmin
	    \\ \\
          \displaystyle
             -\frac{q^2}{4\pi\varepsilon_0 r^3}\vecr, & \rmin  < r \le R_I
	    \\ \\
          \displaystyle
	    \mathbf{0}, & r > R_I.
	\end{array}\right.
    \end{equation}    
    For $r\le R_I$, ion-ion and electron-electron interactions are
    considered as purely Coulombian, i.e.
    $V_{ii}(r)=V_{ee}(r)=\frac{q^2}{4\pi\varepsilon_0 r}$.
    For $r>R_I$, $V_{ii}(r)=V_{ee}(r)=0$.

\subsection{\label{ionrecsim}Ionization-recombination mechanism}

    As discussed in Sec.~\ref{intro}, several works pointed out the treatment
    of electron trapping by a charged ion as a delicate issue when performing
    MD simulations. Here we define a criterion model to deal with such
    scenario and therefore enable a quantitative control of
    ionization-recombination mechanism. Thus, in our simulation model, an
    electron is considered to be \textit{trapped} by an ion when \textit{(a)}
    their mutual distance is less than the
    characteristic atomic size, $\rmin$, and \textit{(b)} the total energy
    of the pair measured in the center-of-mass reference frame becomes
    negative. For such calculation, we only take into account the potential
    energy associated to the corresponding electron-ion pair,
    which is by far the dominant contribution.  When this criterion is
    satisfied, the electron-ion pair is considered as a recombined ion and
    the electron counts as a bound (trapped) one.  Later, due to the
    interaction with remaining particles, the above-mentioned criterion
    conditions may not be satisfied anymore, which is interpreted as an
    ionization, with the electron then returning to the free electron pool.
    Within a classical perspective, this dynamic ionization-recombination
    process is taken into account throughout the simulation run, thus
    allowing neutral pairs to be natural constituents of the final
    equilibrium state. Following this scheme, plasma ionization degree,
    $\alpha$, can be computed at each time instant.

\subsection{\label{pardyn}Particle dynamics and simulation units}

   Particle dynamics follows the Newton's second law,
    \begin{equation}\label{meq}
	m_k\frac{\dif^2\vecr_k}{\dif t^2} =
	  \sum_{k^\prime \atop k\neq k^\prime}\vF_{kk^\prime},
    \end{equation}
    where $m_k$ denotes the mass of the $k$-th particle and $\vecr_k$
    gives its position within the cell. 
    
    From a numerical perspective, in order to accurately solve the system
    of motion equations, proper length $r_p$ and time $t_p$ units have to
    be chosen. Thus, we can write  
    \begin{equation}
	\vecr_k = r_p\,\vecx_k,\quad t = t_p\,\tau,
    \end{equation}
    where $\vecx_k$ and $\tau$ are dimensionless quantities taking values
    of the order of $1$ in the numerical calculation. In terms of $r_p$
    and $t_p$, electron equations of motion can be rewritten as 
    \begin{equation}\label{elecmeq}
	\frac{\dif^2\vecx_k}{\dif\tau^2} = \Gamma_E\,\vF^{(e)}_k,
    \end{equation}
    with
    \begin{equation}\label{gammae}
	\Gamma_E = \frac{1}{m_e}\frac{q^2}{4\pi\varepsilon_0}
		\frac{t_p^2}{r_p^3}.
    \end{equation}
    In Eq.~(\ref{elecmeq}), $\vF^{(e)}_k$ denotes the total force on the
    \textit{k}-th electron measured in units of
    ${q^2}/{4\pi\varepsilon_0r_p^2}$, i.e. the \textit{simulation force
    unit}. A similar expression is found for ions, in which $m_e$ is
    replaced by $m_i$. 

    The average distance between free electrons, $r_0$, might be a
    tempting choice for the simulation length unit $r_p$. However, in the
    course of the simulation some electrons will be trapped by ions, so
    that the exact number of free electrons will be known only when
    achieved the equilibrium state. Therefore, in the simulation, $r_0$ is
    a priori unknown. We then choose for the length unit, $r_p$, the
    average distance between electrons --either free or bound-- within the
    simulation box, i.e.
    \begin{equation}\label{rpdef}
	\frac{L}{r_p} = \left(\frac{4}{3}\pi n_p\right)^{1/3}.
    \end{equation}
    If $n_e$ denotes the number of free electrons within the simulation
    box and $\alpha$ is the \textit{ionization degree}, then 
    \begin{equation}\label{nenp}
	n_e = \alpha n_p, \quad \textrm{with\ } 0\leq\alpha\leq1.
    \end{equation}
    According to Eqs.~(\ref{rpdef}) and (\ref{nenp}) and definition of
    $r_0$, we find
    \begin{equation}\label{rpr0}
	\frac{r_0}{r_p} = \alpha^{-1/3},
    \end{equation}
    i.e. the simulation length unit results a fraction of the free electron
    average distance.

    Similarly, the simulation time unit is defined in terms of the plasma
    characteristic time, $t_0$, in such a way that
    \begin{equation}\label{tpt0}
	\frac{t_0}{t_p} = \beta.
    \end{equation}
    By substituting Eqs.~(\ref{rpr0}) and (\ref{tpt0}) into
    Eq.~(\ref{gammae}), we find 
    \begin{equation}\label{gamegam}
	\Gamma_E = \frac{1}{\alpha\beta^2}\frac{1}{m_e}
	    \frac{q^2}{4\pi\varepsilon_0}\frac{t_0^2}{r_0^3}
	    = \frac{1}{\alpha\beta^2}\Gamma.
    \end{equation}
    Eq.~(\ref{gamegam}) suggests to take $\beta=\alpha^{-1/2}$, so that
    the numerical parameter $\Gamma_E$ matches the simulated \textit{target}
    plasma coupling parameter $\Gamma$, i.e. when launching a simulation,
    one intends to investigate a plasma at certain \textit{target} conditions
    $N_e$ and $T$, which leads to the \textit{target} coupling parameter
    $\Gamma$. However, as discussed above, simulation equilibrium state is a
    priori unknown, so the resulting coupling parameter from the computational
    experiment, $\Gamma_{exp}$, may differ from the originally intended one,
    $\Gamma$. We note in passing that Eq.~(\ref{gamegam}) sets the way to
    define the electron charge value in the numerical calculation. Finally,
    Eq.~(\ref{tpt0}) results
    \begin{equation}\label{tpt02}
	\frac{t_0}{t_p} = \alpha^{-1/2}.
    \end{equation}

    This careful choice of simulation units enables our simulation technique
    to properly deal with the disparate length and time scales that arise in
    the problem of plasma particle dynamics.

    The system of motion equations, Eq.(\ref{meq}), is solved using the
    Verlet's algorithm~\cite{verlet1967,verlet1968} with a certain a time step
    $\Delta\tau$. At each time step, potential, kinetic and total energy of
    the system can be easily calculated. Since the system is conservative,
    total energy must keep constant as time evolves. To numerically satisfy
    this requirement and find a correct solution of dynamical equations, a
    proper $\Delta \tau$ must be used. In this work, $\Delta \tau$ is chosen
    so that the average of energy numerical fluctuations is equal to zero
    during the simulation. This is a demanding requirement that typically
    leads to a short $\Delta\tau$ value and, accordingly, to a high
    computational cost. Nevertheless, as discussed in
    Sec.~\ref{modelsimresults}, we benefit from the fact that no
    \textit{numerical heating}~\cite{hockneybook,ueda1994,rambo1997,
    evstatiev2013} is observed, thus avoiding the use of \textit{numerical
    thermostats}~\cite{hunenberger2005}.  

    Going through Eqs.~(\ref{pot}-\ref{tpt02}), one can see that simulation
    depends on only two independent physical parameters, i.e. $V_i$ and
    $\Gamma_E$, and the number of particles, $n_p$. This makes simulation to
    exhibit interesting scaling properties. Results obtained in
    \textit{simulation units} can be expressed in absolute physical units and
    understanding of such unit conversion is very important for the
    corresponding physical interpretation. For instance, working in simulation
    units, let us suppose a simulation launched with $V_i=4.75$
    ($\equiv4.75\ene_0$, being $\ene_0$ the simulation energy unit),
    $\Gamma_E=0.116$ and $n_p=255$. Also, suppose that, at equilibrium, the
    simulation gives $\alpha=0.53$ and we \textit{numerically measure} a
    kinetic energy per particle of $\ene_k=0.50$ ($\equiv0.50\ene_0$). Now
    if we are actually interested in simulating a hydrogen plasma, then
    $V_i=13.6$~eV. This fact sets the simulation energy unit, i.e.
    $\ene_0=2.86$~eV.  Since
    $\ene_k=\frac{1}{2}m_e\langle v^2\rangle=\frac{3}{2}kT$, from the kinetic
    energy measurement we have the plasma temperature at equilibrium, i.e.
    $kT=0.95$~eV. Using Eq.~(\ref{asize}), in physical units, the
    characteristic atomic size results $a=1.59$~\AA. Considering
    Eqs.~(\ref{asize}) and (\ref{gammae}), and taking into account that in
    simulation units $m_e=1$, $r_p=1$, and $t_p=1$, in terms of input
    parameters then we have
    $a=\frac{3}{2}\frac{\Gamma_E}{V_i}=0.0366$($\equiv0.0366r_p$). Thus, the
    simulation length unit equivalent is determined, i.e. $r_p=43.39$~\AA.
    Using Eq.~(\ref{rpdef}), one may calculate the size of simulation box,
    i.e. $L\approx443$~\AA. Finally, from Eq.~(\ref{rpr0}) and definition of
    $r_0$, we obtain the plasma electron density at equilibrium, i.e.
    $N_e=1.55\times10^{24}$~m$^{-3}$.

\subsection{\label{initcond}Setup of initial conditions}

    In order to initialize the system of motion equations, initial positions
    and velocities of all particles in the simulation must be specified. Ion
    positions are drawn following a uniform distribution within the volume of
    the simulation box. Around each ion, one electron
    is randomly placed in
    a spherical surface of a certain radius. For such configuration, the
    potential energy is mainly given by the sum of 
    binding energy of ion-electron pairs,
    since the contribution from the remaining particles
    is almost negligible. Thus, a proper choice of the ion-electron distance
    allows to easily set the initial potential energy to the desired value.
    Particle velocities are randomly set according to a Maxwellian
    distribution characterized by a certain temperature and the corresponding
    mass for each type of particle.

    Initial configuration determines the initial potential and kinetic
    energy and, therefore, the total energy of the system.
    Also, some of the ion-electron pairs may satisfy the criterion described
    in Sec.~\ref{ionrecsim} to be considered as a recombined-ion, so that in
    general initial ionization degree does not correspond to the fully-ionized
    state. In other words, the ionization degree is not set by means of any
    fine-tuning of initial conditions, such initial value is the one that
    naturally results from the initial configuration.
    
    For $t>0$, the system will evolve undergoing multiple ionizations and
    recombinations and an exchange between potential and kinetic energy will
    occur until the system reaches the corresponding equilibrium state and
    ionization balance. Typically, at $t\sim0$ a sudden exchange between
    kinetic and potential energy takes place, which is interpreted as a
    natural \textit{readjustment} of the initial configuration. Initially,
    particle velocities are assigned following a Maxwellian distribution,
    so that initial kinetic energy is already \textit{well} distributed.
    However, this is not the case of potential energy.  At $t\sim0$ all
    particles have a potential energy value very similar to the one
    resulting from the initial draw. This leads to an initial potential
    energy distribution which resembles to a Dirac $\delta$-function and that
    is certainly far from the one to be reached at equilibrium ---see
    Fig.~\ref{potener_dist}---. As a consequence, early in time, such
    configuration will evolve very quickly. Just a small collective movement
    of the order of $r_p$ is sufficient to produce a significant exchange
    between kinetic and potential energy. This occurs in a time lapse of the
    order of $t_p$, so definitely the initial energy exchange will be observed
    as a sudden event compared to system evolution typical time scale.

    In the entire
    process, the total energy will remain ---within numerical fluctuations---
    constant. We recall that the generated time-histories of physical
    quantities will be useful for statistical purposes only after equilibrium
    has been reached. Setup of initial conditions described here permits to
    easily manage the balance between initial potential and kinetic energy,
    which with the guidance provided by the equilibrium model developed in
    Sec.~\ref{eqmodel} eventually represents a way to speed up the simulation
    to reach the equilibrium state without any artificial numerical
    adjustments. In particular, we recall that no thermostat algorithm has
    been used.

\subsection{Computational resources and details}

    Simulations were run in parallel in a computer cluster equipped with a
    total of 52 graphics processing units (GPU). All computer programs
    referred in this work have been coded  from scratch using C$^{++}$ and
    CUDA\textsuperscript{\textregistered}. No commercial software neither
    public domain code was used. We actually developed two different codes,
    one to be run on CPU (sequentially) and the other on GPU (in parallel).
    This allowed us an easier debugging of our programs, since starting
    from the same initial conditions, both versions must lead to the same
    results. Running on CPU was approximately 30 times slower than doing it
    on GPU, so CPU version was obviously used only for debugging purposes.

    When working on a GPU, CUDA\textsuperscript{\textregistered} programming
    model distinguishes between \textit{threads} ---the smallest execution
    unit--- and \textit{blocks} ---a group of threads---. Also,
    CUDA\textsuperscript{\textregistered} memory hierarchy consists of
    multiple memory spaces. For instance, each thread has its own private
    local memory and each block has shared memory visible to all block threads
    with the same lifetime as the block. In our simulation code, each block
    deals with one simulation box, i.e. one plasma sample, and each thread
    within a block is responsible for one single particle. This way all
    threads run exactly the same piece of code but using different numerical
    data. Particle locations and interactions between them are saved in the
    block shared memory, whereas thread local memory stores the velocity of
    the associated particle. Calculation of interactions is performed in three
    steps. In the first one code computes the repulsive force between electrons,
    the repulsive force between ions in the second step and the attractive
    force between ions and electrons in the last one. For calculation of
    repulsive interactions a \textit{do-loop} is launched for
    $i = 1,\;\dots,(n_p-1)/2$, with $n_p$ always being an odd integer ---we
    recall that $n_p$ denotes the number of electrons (and ions) in the
    simulation---. In the $i$-th loop iteration, every $j$-th thread computes
    the interaction between the $j$-th particle and the one with index
    $(j+i)\; \mathrm{mod}\; n_p$ ---coded as {\tt (j+i)\% np} in C$^{++}$---.
    Force corresponding to the symmetric configuration is obtained according
    to Newton's third law and for that reason only $(n_p-1)/2$ iterations are
    needed. On every loop iteration, computed force is stored in the block
    shared memory. This action will never cause a memory conflict because
    there is not one thread-pair working with the same interaction. At the
    end of every loop iteration, all threads are synchronized thus preventing
    any memory conflict during next iteration. A similar algorithm is used for
    attractive interactions. For these algorithms cache is loaded only once
    and remains unchanged for the specified number of time steps.
    
    The major limitation of the code comes from the cache size, which restricts
    the maximum number of particles than can be simulated. All calculations
    presented here were performed this way, which in turn represents the
    fastest, well-tested and reliable version of the simulation code. In order
    to increase the total number of particles operations must be distributed
    among different blocks, thus breaking the one-by-one correspondence between
    plasma samples and blocks. We also developed such a version, which is slower
    due to required synchronization between different blocks and consequently
    more prone to cache faults occurrence.

    The entire set of calculations ---not all of them shown here--- and
    analyses performed to build up the present research spanned over six
    wall-clock time months. In total, we carried out 1488 independent
    simulations that took a range of computational times from 1 to 81 days
    (typically 5 days)
    depending on physical conditions of simulated plasma.

\section{Analytical statistical equilibrium model}\label{eqmodel}

    Computer simulations are often considered as idealized experiments providing
    a unique testbed for validation of theoretical models. In this regard,
    assessment of simulation reliability becomes a critical task that not
    always has received the attention it deserves. Here we have developed an
    analytical model to describe the equilibrium state of a hydrogen plasma
    and thus checking the reliability and validity of the computer
    simulations. This model does not aim to describe the behavior of a
    \textit{real} plasma but to mimic the physical conditions in which
    simulation takes place and properly describe the corresponding
    statistics.

    Dissipative radiative processes are not taken into account in the
    simulation, so the ionization balance appears as a result of collisional
    ionization and recombination processes. In this context, population
    kinetics is ruled by the well-known Saha equation~\cite{saha1920},
    \begin{equation}\label{eqSaha}
	\frac{n_e n_i}{n_n} = \frac{Z_e(T)Z_i(T)}{Z_n(T)},
    \end{equation}
    wherein $n_e$, $n_i$ and $n_n$ are the number of free electrons, ions
    and neutral atoms ---in the simulation a neutral atom consists of a
    bound electron-proton pair--- in the plasma, respectively, and
    $Z_e(T)$, $Z_i(T)$ and $Z_n(T)$ are the corresponding classical partition
    functions at equilibrium temperature $T$. 

    Particle dynamics is ruled by laws of classical mechanics and
    accordingly it will show classical statistical properties. Hence, 
    the free electron partition function is given by
    \begin{eqnarray} \label{Ze}
	Z_e(T) &=& \int_V\,\dif^3\vecr\;\int\,\dif^3
		\vecp\exp\left(-\frac{p^2}{2m_ekT}\right)
		\nonumber\\
	       &=& V\left(2\pi m_e kT\right)^{3/2}.
    \end{eqnarray}
    Similarly, for ions we have
    \begin{equation}\label{Zi}
	Z_i(T) = V\left(2\pi m_i kT\right)^{3/2}.
    \end{equation}
    Lastly, neutral atoms partition function is obtained as
    \begin{equation}\label{Zn}
	Z_{n}(T) = Z_{n\;{\rm trans}}(T)\,Z_{n\;{\rm int}}(T),
    \end{equation}
    \begin{widetext}
    with	
    \begin{equation}
	\label{Zntrans}
	Z_{n\;{\rm trans}}(T) = \int_V\dif^3\vecr\int\dif^3
		\vecp\exp\left(-\frac{p^2}{2m_nkT}\right)
                = V\left(2\pi m_n\,kT\right)^{3/2},
    \end{equation}                                 
    \begin{eqnarray}
	\label{Znint}
	Z_{n\;{\rm int}}(T) &=& \int_{\ene < 0}\dif^3\vecr
		\int\dif^3\vecp\exp\left\{
		-\frac{1}{kT}\left[
		\frac{p^2}{2\mu}+V_{ie}(r)\right]\right\}
		\nonumber\\
                &=& \left(
		    \frac{3\pi kT\,\rmin^2}{V_i}\right)^{3/2}
		    \left(2\pi\mu kT\right)^{3/2}\,
		    {\rm e}^{V_i/kT}
		\left\{
		    1-{\rm e}^{-V_i/kT} \left[
		       1 + \left(
		       \frac{V_i}{kT}\right)
		       + \frac{1}{2}\left(\frac{
		       V_i}{kT}\right)^2\right]\right\}.
    \end{eqnarray}
    \end{widetext}
    Here, $V$ stands for the plasma volume, $m_n= m_i+m_e$, and
    $\mu = m_i m_e/m_n$ is the ion-electron reduced mass.
    $Z_{n\;{\rm trans}}(T)$ denotes the translational partition function,
    i.e. resulting from the movement (translation) of the center of mass
    and $Z_{n\;{\rm int}}(T)$ is the internal partition function, which
    accounts for internal degrees of freedom. In Eq.~(\ref{Znint}) the
    integration domain is limited to the phase space region satisfiying 
    \begin{equation}\label{eneint}
    	\ene(r,p) = \frac{p^2}{2\mu} + V_{ie}(r) < 0,
    \end{equation}
    as happens in a bound system. We also note that the choice of the
    quadratic behaviour for $V_{ie}(r)$ at short distances, Eq.~(\ref{pot}),
    leads to an analytical solution for the coordinates integral in
    Eq.~(\ref{Znint}).  

    \begin{widetext}
    Thus, our \textit{classical} Saha equation results 
    \begin{equation}\label{eqSahaC}
	\frac{N_e\,N_i}{N_n} = 
		\left(\frac{V_i}{kT}\frac{1}{3\pi\rmin^2}\right)^{3/2}
		\frac{\displaystyle{\rm e}^{-V_i/kT}}{\displaystyle
		  1-{\rm e}^{-V_i/kT}
		  \left[
		    1 + \left(\frac{V_i}{kT}\right)
		    + \frac{1}{2}\left(\frac{V_i}{kT}\right)^2
		  \right]},
    \end{equation}
    with $N_x = n_x/V$ and $x\equiv e,\ i,\ n$.

    With $N_p=N_i+N_n$, in terms of the ionization degree, $\alpha$,
    we have $N_e=N_i=\alpha N_p$ and $N_n=(1-\alpha)N_p$. Then,
    Eq.(\ref{eqSahaC}) becomes
    \begin{equation}\label{eqAlpha2}
	\frac{\alpha^2}{1-\alpha} = K(T)
    \end{equation}
    being
    \begin{equation}\label{Keqmodel}
	\hspace{-3em}
	K(T) = \frac{4}{9\sqrt{3\pi}}\left(
                \frac{r_p}{\rmin}\right)^3
                \,\left(\frac{V_i}{kT}\right)^{3/2}
                \frac{\displaystyle{\rm e}^{-V_i/kT}}{\displaystyle
                  1-{\rm e}^{-V_i/kT}
                  \left[
                    1 + \left(\frac{V_i}{kT}\right)
                    + \frac{1}{2}\left(\frac{V_i}{kT}\right)^2
                  \right]},
    \end{equation}
    \end{widetext}
    where $N_p=1/(\frac{4}{3}\pi r_p^3)$. As seen from
    Eq.~(\ref{Keqmodel}), plasma ionization balance does not depend on
    ion mass. Thus, we finally obtain a compact analytical formula for
    the plasma ionization degree,
    \begin{equation}\label{alphaeqmodel}
	\alpha = \frac{K(T)}{2}\left[
                \sqrt{1+\frac{4}{K(T)}} - 1\right].
    \end{equation}
    As shown in detail in Sec.~\ref{modelsimresults}, last equation will
    be very useful to analyze the results provided by the numerical
    simulations.

\subsection{Equilibrium curve}

    Besides providing the plasma ionization degree, the analytical model
    allows us to build an \textit{equilibrium curve}, i.e.  
    a potential energy vs kinetic energy plot that shows how the total
    energy distributes on each equilibrium state. 

    In this context, the average kinetic energy is simply given by
    \begin{equation}\label{ekeqmodel}
	{\cal E}_k= \frac{3}{2}kT,     
    \end{equation}
    and the average potential energy of free particles is considered to be
    equal to zero. The potential energy of an ion-electron pair in a
    neutral atom is given by its binding energy, since this is by far the
    most important contribution and the ones from remaining particles can
    be neglected. Hence, the potential energy per bound particle is written
    as
    \begin{equation}\label{epbeqmodel}
	{\cal E}_{pb} = \frac{1}{2}\left(-V_i+\frac{3}{2}kT\right),     
    \end{equation}
    as corresponds to a parabolic potential. The $\frac{1}{2}$ common
    factor distributes the binding energy between the proton and the
    electron in the pair. The potential energy per particle is given by 
    \begin{equation}\label{epeqmodel}
	{\cal E}_{p} = \frac{1-\alpha}{2}\left(-V_i+\frac{3}{2}kT\right).
    \end{equation}
    
\section{\label{results}Results}

    The investigation addressed in this work relies on a significant
    number of MD simulations of both electron-positron and electron-proton
    plasmas. Comprehensive studies to check the consistency of the simulation
    code and physical behaviour with respect to simulation parameters like
    the potential-energy well depth or the coupling parameter were performed.
    For the sake of clarity, the entire set of calculations are summarized in
    Table~\ref{symtab}, with indications of input parameter values and other
    simulations details, such us the chosen time step and kinetic and
    potential energy initial conditions. Obviously, there is no point to
    show all the collected calculations here. Representative results have
    been properly chosen according to the aims of this work: the study of the
    equilibration process and the computational validity assessment of our
    simulation model. These topics and results are discussed in the following
    Secs.~\ref{eqtime} and \ref{modelsimresults}. 

\begin{table*}[t]
  \caption{\label{symtab}%
   Summary of the entire set of calculations performed to build up the
   present work. Specifications of the electron-positron and electron-proton
   simulated plasmas are given. These include:
   (a) the numerical coupling parameter, $\Gamma_E$;
   (b) the ionization energy, $V_i/{\cal E}_0$ ;
   (c) the number of electrons, $n_p$, included in the simulation
        ---which equals the number of positrons or protons---;
	the ranges of initial
   (d) kinetic and
   (e) potential energy ---examples of specific initial values are indicated
	as open circles (green) associated to $t=0$ in
	Fig.~\ref{modsim_eqcurve}---;
   (f) the range of time step values used to integrate the motion equations,
        $\Delta t/t_p$;
   (g) the range of values for the number of time steps reached in the
	simulation and
   (h) the number of simulation runs associated to the specified input
	parameter values ---indicated as $c\times g$, where $c$ is the
	number of cases and $g$ is the number of plasma samples sharing
	the same initial physical conditions---.
  }
  \begin{ruledtabular}
  \begin{tabular}{cccccccccc}
    \\
    \multicolumn{10}{l}{\textbf{Type of plasma:} Electron-positron}\\
    \hline
     \multicolumn{3}{c}{Input parameters} & &
     \multicolumn{2}{c}{Initial conditions} & &
     Time step & \# $\times 10^6$ steps & \# Simulation \\
    \cline{1-3} \cline{5-6} \cline{8-9} 
     $\Gamma_E$ & $V_i/{\cal E}_0$ & $n_p$ & &
     ${\cal E}_k/{\cal E}_0$ range & ${\cal E}_p/{\cal E}_0$ range & &
     $\Delta t/10^{-4}t_p$ & $t_\mathrm{total}/10^{6}\Delta t$ &
     runs \\
    \cline{1-3} \cline{5-6} \cline{8-9} \cline{10-10}
0.1161 & 3.00 & 255 & & $\left[0.00,3.00\right]$ & $\left[-1.25,-0.25\right]$ & & $4$                & $\left[140,300\right]$  & $18\times 8$ \\
0.1161 & 4.75 & 255 & & $\left[0.00,3.00\right]$ & $\left[-2.25,-0.25\right]$ & & $\left[1,5\right]$ & $\left[36,300\right]$   & $53\times 8$ \\
0.1161 & 5.50 & 255 & & $\left[0.00,3.00\right]$ & $\left[-2.70,-0.25\right]$ & & $\left[2,4\right]$ & $\left[133,300\right]$  & $20\times 8$ \\
0.1161 & 6.80 & 255 & & $\left[0.00,3.00\right]$ & $\left[-3.25,-0.25\right]$ & & $\left[1,5\right]$ & $\left[67,237\right]$   & $33\times 8$ \\
0.0417 & 4.75 & 425 & & $\left[0.00,3.00\right]$ & $\left[-2.00,-0.25\right]$ & & $2$                & $\left[15,1445\right]$  & $15\times 8$ \\
0.0417 & 6.80 & 425 & & $\left[0.00,3.00\right]$ & $\left[-3.25,-0.25\right]$ & & $5$                & $\left[487,1309\right]$ & $33\times 8$ \\
    \hline
    \\
    \multicolumn{10}{l}{\textbf{Type of plasma:} Electron-proton}\\
    \hline
     \multicolumn{3}{c}{Input parameters} & &
     \multicolumn{2}{c}{Initial conditions} & &
     Time step & \# $\times 10^6$ steps & \# Simulation \\
    \cline{1-3} \cline{5-6} \cline{8-9} 
     $\Gamma_E$ & $V_i/{\cal E}_0$ & $n_p$ & &
     ${\cal E}_k/{\cal E}_0$ range & ${\cal E}_p/{\cal E}_0$ range & &
     $\Delta t/10^{-4}t_p$ & $t_\mathrm{total}/10^{6}\Delta t$ &
     runs \\
    \cline{1-3} \cline{5-6} \cline{8-9} \cline{10-10}
0.1161 & 4.75 & 255 & & $\left[0.00,3.00\right]$ & $\left[-2.00,-0.25\right]$ & & $\left[0.5,5\right]$ & $\left[40,1886\right]$ & $14\times 8$ \\
  \end{tabular}
  \end{ruledtabular}
\end{table*}

\subsection{\label{eqtime}Equilibration process}

    In a MD simulation the statistical sampling of relevant physical
    quantities and processes is only meaningful once particle dynamics
    becomes stationary and the system thus reaches the equilibrium
    state. This means that a MD calculation has to go through an initial
    equilibration or relaxation stage, which is by itself useless to get
    relevant physical information, but in turn necessary to drive the
    system to the stationary stage from which the statistical sampling
    can be safely performed. From a computational point of view, time
    needed to reach the equilibrium state in a simulation run is
    substantial. Typically, for MD simulations of
    \textit{positron-electron} plasmas, equilibration time easily hits
    a few thousands of simulation time units, i.e. $\sim10^3\,t_p$, which
    is in agreement with the results obtained in Ref.~\cite{stambulchik2007}.
    With the choice of an integration time step of the order of
    $10^{-4}\,t_p$ for solving the system of motion equations, reaching
    the stationary stage therefore requires several millions of time
    steps. In this regard, caution must be taken to do not prematurely
    terminate simulation runs, which might lead to an inaccurate
    statistics of physical quantities.

\begin{figure}[t]
    \begin{center}
    \includegraphics[width=\columnwidth]{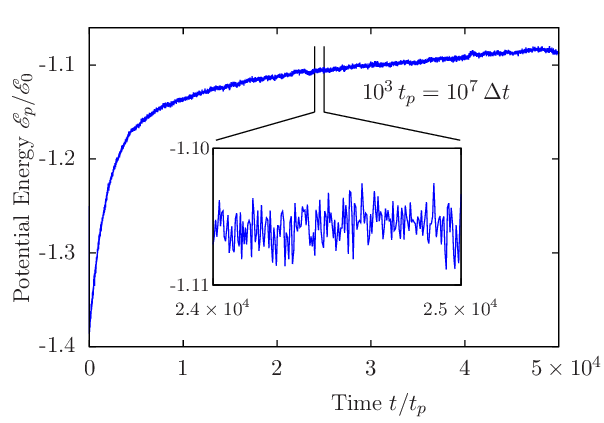}
    \caption{\label{FIG08}
    	Time evolution of potential energy per particle in a simulation
	of a electron-proton plasma ($m_p=1836m_e$). Throughout the
	thermalization process, the inspection of a time interval
	including tens of millions steps may result too short to
	securely identify that equilibrium has been reached.}
    \end{center}
\end{figure}

    Equilibration time further increases with mass difference between
    plasma constituents ---electrons and heavier ions, for instance---,
    since very different time scales appear then involved in motion
    equations. Figure~\ref{FIG08} illustrates the delicate and slow
    process of plasma equilibration. In this case a \textit{proton-electron}
    plasma ---i.e. hydrogen plasma--- has been simulated. Figure shows
    the time evolution of potential energy per particle. From the inset
    plot it is clear that when looking at a time interval including
    \textit{only} tens of millions time steps the change in energy appears
    hidden by numerical fluctuations. This may lead to wrongly think that
    the equilibrium has been achieved and thus prematurely terminate the
    simulation run. In the example, the true equilibrium state is still far
    from being reached. In this regard, it is certainly helpful to have a
    model such as the one described in Sec.~\ref{eqmodel} to provide
    guidance about the equilibrium point
    $ ({\cal E} _k^{eq}, {\cal E} _p^{eq}) $ and undoubtedly identify the
    thermalization of the simulated plasma. We note in passing that, in
    this work, thermalization ---i.e. equilibrium state--- is considered
    to occur when statistical distributions of all physical quantities
    become stationary.  

\begin{figure}[t]
    \includegraphics[width=\columnwidth]{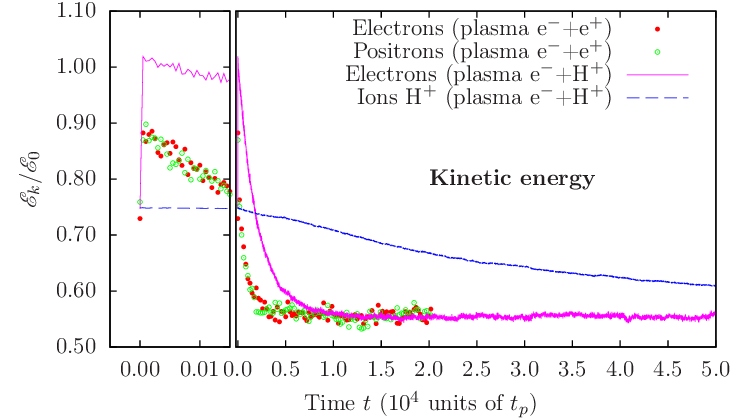}
    \caption{\label{ekevol_comp}
    	Comparison of equilibration process between
	\textit{positron-electron} and \textit{proton-electron} plasmas.
	Here we plot the time history of the kinetic energy per particle.
	The characteristic sudden evolution early in time from the initial
	configuration has been zoomed in on the left box.}
\end{figure}

\begin{figure}[t]
    \includegraphics[width=\columnwidth]{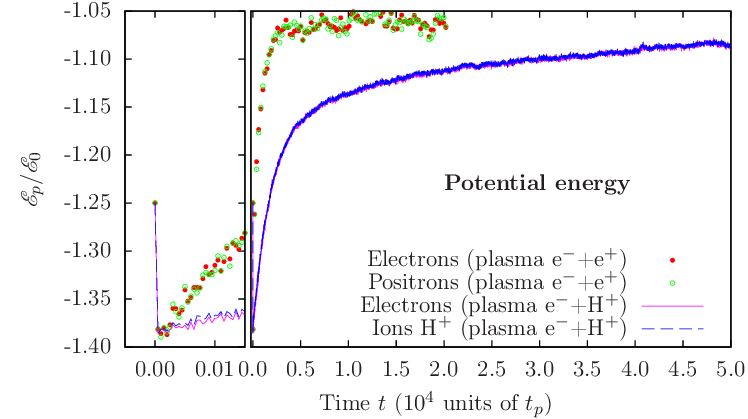}
    \caption{\label{epevol_comp}
    	Time history of potential energy per particle for the same cases
	shown in Fig.~\ref{ekevol_comp}.}
\end{figure}

\begin{figure}[t]
    \includegraphics[width=\columnwidth]{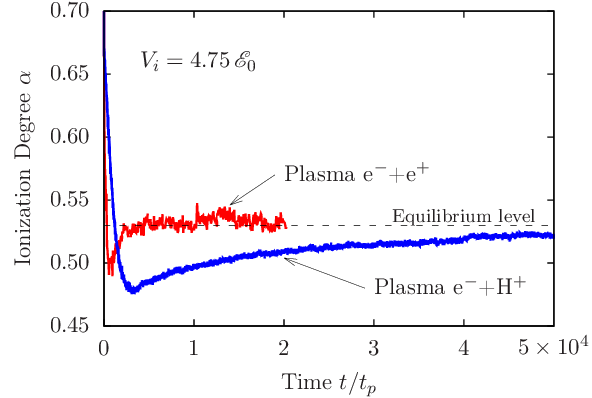}
    \caption{\label{ionevol_comp}
    	Time history of plasma ionization degree for the same cases shown
	in Figs.~\ref{ekevol_comp} and~\ref{epevol_comp}.}
\end{figure}

    Equilibration process is much faster in a \textit{positron-electron}
    plasma than in a \textit{proton-electron} case. For comparison, time
    evolution of kinetic and potential energy per particle, and plasma
    ionization degree are shown in
    Figs.~\ref{ekevol_comp},~\ref{epevol_comp}~and~\ref{ionevol_comp},
    respectively. While ions take more time to thermalize, electron
    kinetic energy rapidly reaches the stationary state even for a hydrogen
    plasma. A slower path to equilibrium is observed for the case of
    potential energy, i.e. particle spatial distribution takes longer to
    achieve an equilibrium configuration. This is particularly important
    because our simulations are ultimately aimed to study and characterize
    the local electric field properties, and obviously the corresponding
    dynamics and statistics are ruled by particle spatial arrangement.  

    Keeping in mind the criterion adopted in the simulation to model the
    ionization-recombination mechanism ---described in
    Sec.~\ref{ionrecsim}---, plasma ionization degree can be computed at
    each time instant throughout the system evolution. An example is shown
    in Fig.~\ref{ionevol_comp}.
    Compared to kinetic and potential energy, the ionization degree shows
    the slowest approach to equilibrium. This is because, as suggested
    before, from a classical perspective the ionization balance equilibration
    process is basically ruled by the less-frequent three-body processes.
    Thus, even when a small change in the kinetic energy of a given electron
    or in the potential energy associated to a positron-electron (or
    ion-electron) pair have a negligible influence on the corresponding
    average values, such a small change may determine the difference for an
    electron to be considered either as a bound or a free one, which indeed
    has a greater impact on the calculation of the ionization degree.
    Stationarity assessment of this parameter is crucial, since it determines
    the free electron density value, which is a key quantity for the
    statistical analysis of the local electric field. 

    As Eqs.~(\ref{Keqmodel}), (\ref{alphaeqmodel}) and (\ref{epeqmodel})
    suggest, neither energy distribution among particles nor partition
    between kinetic and potential energy depend on particle mass. We have
    both numerically confirmed this fact and taken advantage of it to speed
    up the simulation of hydrogen plasmas. We launch the calculations using
    positrons and once the equilibirum is reached, positron mass is replaced
    by proton mass and velocity moduli are modified accordingly to keep the
    right kinetic energy values. Thenceforth, taking the particle spatial
    distribution at the switch time, the simulation proceeds with updated
    masses and velocities. 

\begin{figure}[t]
  \begin{center}
    \includegraphics[width=\columnwidth]{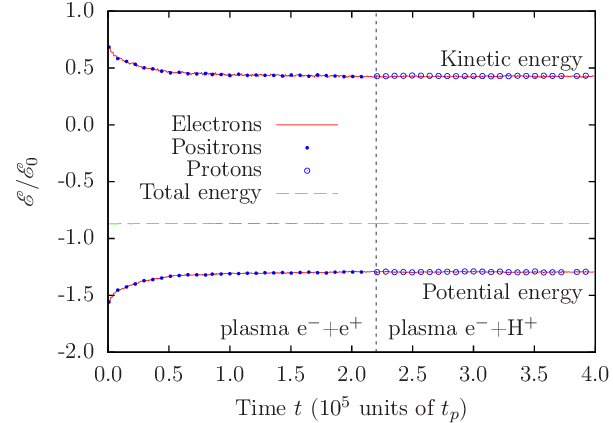}
    \caption{\label{posprorepl1}
	Time histories of kinetic, potential and total energy per
	particle for a simulation that begins as a electron-positron
	plasma and converts to a hydongen plasma. The positron-by-proton
	replacement occurs at $t_c = 2.2\times10^{5}t_p$. Plot illustrates
	the technique to speed up the process for achieving the
	equilibrium in a hydrogen plasma.}
  \end{center}
\end{figure}

\begin{figure}[t]
  \begin{center}
    \includegraphics[width=\columnwidth]{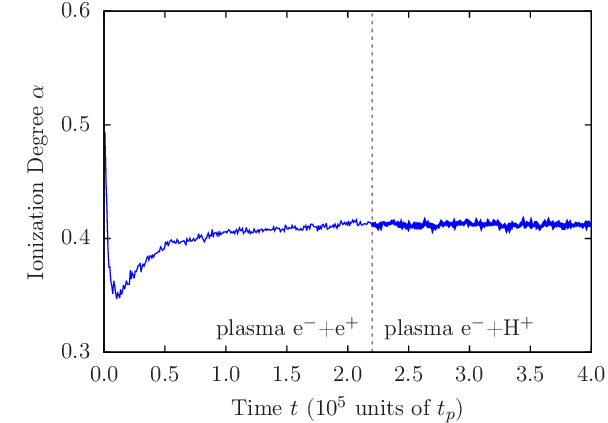}
    \caption{\label{posprorepl2}
	Time history of plasma ionization degree for the same case shown
	in Fig.~\ref{posprorepl1}.}  
  \end{center}
\end{figure}

\begin{figure}[t]
  \begin{center}
    \includegraphics[width=\columnwidth]{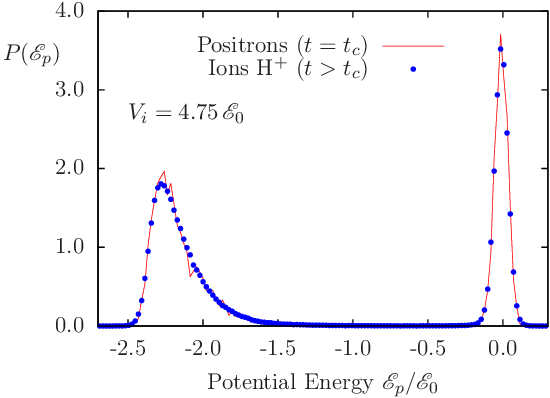}
    \caption{\label{posprorepl3}
    	Statistical distribution of potential energy per particle. Results
	are shown for (a) positrons (in a electron-positron plasma) at the
	time $t_c$ of positron-by-proton replacement and (b) protons
	(electron-proton plasma) after the switch time ---see the text for
	details---.}
  \end{center}
\end{figure}

    We illustrate this point in Figs.~\ref{posprorepl1}, \ref{posprorepl2}
    and \ref{posprorepl3}.  Fig.~\ref{posprorepl1} shows time histories of
    kinetic, potential and total energy per particle. Evolution of plasma
    ionization degree is plotted in Fig.~\ref{posprorepl2}. When going
    through the positron-proton switch time, we did not observe any
    appreciable difference in either average energy values or plasma
    ionization degree, which show the characteristic steady behavior at
    equilibrium. Also, with the only expected exception of the ion
    velocity distribution, it is seen that statistical distributions of
    the system do not change with the positrons-by-protons (ions)
    replacement. An illustration is given in Fig.~\ref{posprorepl3}.
    For the case of
    positrons, the potential energy statistical distribution shown in the
    figure actually represents the average result over 8 simulation runs with
    the same plasma macroscopical physical conditions sampled at the time
    $t_c$, i.e. right before the replacement. The distribution corresponding
    to hydrogen ions has been obtained using data from the same 8 independent
    simulations but further sampling each simulation every 1000 time units
    throughout a $2.64\times10^5$ units long time interval at equilibrium
    conditions --i.e. result shown in the figure actually sum up the data
    from 2112 different plasma configurations taken from simulation time
    histories within the stationary stage--.
    We also note that the duration of the referred interval after switch
    time is equivalent to $\sim 6000$ times the proton characteristic time,
    which is long enough for each proton to go $\sim 700$ times across the
    simulation box. As seen,
    a good consistency between
    distributions before and after updating mass and velocity of positive
    charged particles was obtained.
%

\subsection{\label{modelsimresults}Comparison with the equilibrium model
    and computational validity assessment}

    Our goal in this section is the computational validity assessment of simulation
    results by comparison with the analytical theoretical model described in
    Sec.~\ref{eqmodel}. Results shown here were obtained from a
    considerable number of positron-electron plasma simulations. As shown in
    Sec.~\ref{eqtime}, the system properties and configuration at equilibrium
    do not depend on particle masses, so we take advantage of this fact and
    use positrons with the only purpose to speed up the simulations and reduce
    the computational time to achieve the equilibrium.  For each simulation,
    positions and velocites of 255 (or 425) positrons and electrons were drawn
    according to the procedure described in Sec.~\ref{initcond}. In order to
    obtain more accurate statistical distributions and average values, for
    each pair $({\cal E}_k,{\cal E}_p)$ of kinetic and potential energies, we
    performed from 8 to 64 completely independent simulations, i.e. initial
    positions and velocities were different, but yielding the same average
    potential and kinetic energy values per particle. 

    As discussed in Sec.~\ref{mdmodel}, numerical algorithms employed in our
    simulation model are robust enough, so that throughout the evolution of
    simulated plasma no external control procedure of total energy value was
    needed. Time step for integration of motion equations was chosen between
    $5\times 10^{-5}$ and $5\times 10^{-4}$ depending on the case, i.e. time
    step value is taken lower the greater the average kinetic energy. As a
    rule of thumb, in a time step a plasma particle travels a distance of the
    order of $10^{-4}$ times the characteristic interparticle distance. No
    numerical heating was observed.

\begin{figure}[h]
  \begin{center}
    \includegraphics[width=\columnwidth]{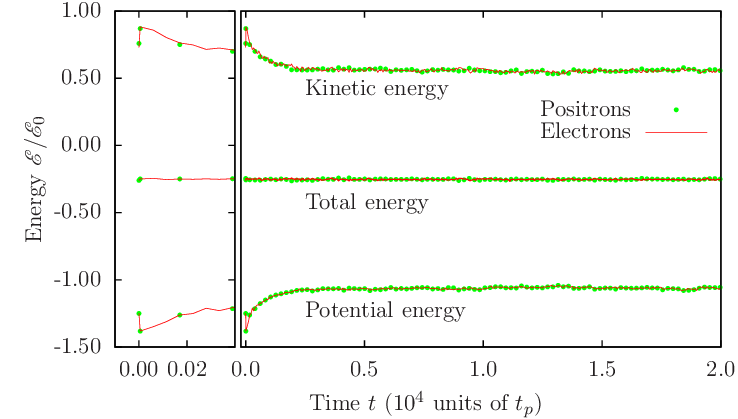}
    \caption{\label{en_evol}
    	Illustration of kinetic, potential and total energy time histories
	in a simulation. For the case shown, average kinetic and potential
	energies in the initial configuration were
	${\cal E}_k=0.74{\cal E}_0$ and ${\cal E}_p=-1.25{\cal E}_0$,
	respectively; and $V_i = 4.75{\cal E}_0$. Results represent the
	average over 8 simulations launched with the same initial average
	kinetic and potential energies.  On the left side, simulation
	earliest stage has been zoomed in. On the right side, the entire
	time history is shown. For the sake of clarity, only 1 every 10
	calculation time steps are displayed in the electron time histories
	and only 1 every 40 in the case of positrons.}
  \end{center}
\end{figure}

    For illustration of equilibration process in the simulation runs shown in
    this section, in Fig.~\ref{en_evol} we plot the evolution of kinetic,
    potential and total energy per particle for a given simulation case.
    Results have been obtained by averaging 8 plasma simulations launched with
    the same initial kinetic and potential energies. Typically, at earliest
    stage of the simulation, a sudden change in both particle positions and
    velocities happens as a consequence of a kinetic and potential energy
    exchange. Then, a slower evolution towards the equilibrium is observed.
    In the case shown a transfer from kinetic to potential energy ocurred,
    although the opposite might happen in other cases. As seen, total energy
    is conserved throughout the entire simulation and only the characteristic
    numerical fluctuations are observed. 
  
\begin{figure}[t]
  \begin{center}
    \includegraphics[width=\columnwidth]{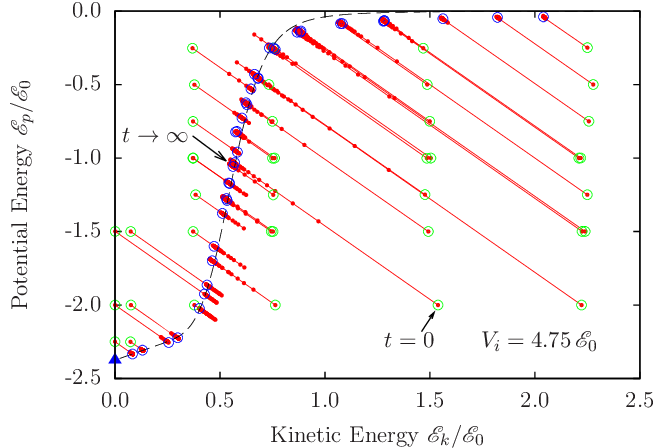}
    \caption{\label{modsim_eqcurve}
    	Comparison between the equilibrium curve predicted by the
	analytical model (dashed line) and the one obtained from
	numerical simulation runs. Path to equilibrium for each
	simulation run ---solid circles--- follows a
	${\cal E}_k+{\cal E}_p = \mathrm{cte}$ trajectory.
	Starting ---i.e. initial configuration--- and final points
	of simulation trajectories are indicated by open circles
	(green and blue, respectively).
	The triangle dot indicates the minimum energy state of
	the system ---i.e. the case of a static distribution of
	ions and bound electrons---. Simulation input parameters
	were $V_i = 4.75{\cal E}_0$ and $\Gamma_E=0.116$.}
  \end{center}
\end{figure}

\begin{figure}[t]
  \begin{center}
    \includegraphics[width=\columnwidth]{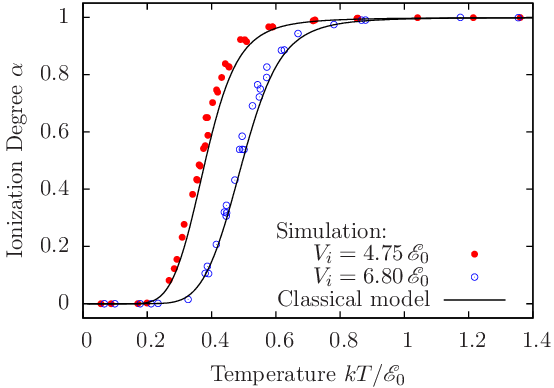}
    \caption{\label{modsim_alpha}
    	Ionization degree as a function of temperature at
	equilibrium. Comparisons between the analytical model and
	simulations are shown for two cases, i.e.
	$V_i=4.75{\cal E}_0$ and $V_i=6.80{\cal E}_0$ (with $\Gamma_E=0.116$).}
  \end{center}
\end{figure}

    In order to illustrate the validity assessment of our MD code, comparisons
    between the analytical equilibrium model and simulation results are
    displayed in Figs.~\ref{modsim_eqcurve}, \ref{modsim_alpha} and
    \ref{modsim_etot}. In particular, Fig.~\ref{modsim_eqcurve} shows the
    equilibrium curve as predicted by the model according to
    Eqs.~(\ref{ekeqmodel}-\ref{epeqmodel}).  Paths to equilibrium in such
    ${\cal E}_p$:${\cal E}_k$ plane are also plotted for a collection of
    40 different simulation cases. Potential and kinetic energy values
    of initial configurations span over an interval broad enough to survey the
    equilibrium curve in a wide temperature range. For each case, the plasma
    approaches the equilibrium by means of an exchange between kinetic and
    potential energy, and always following a ${\cal E}_k+{\cal E}_p=cte$
    trajectory. When simulation starts at a point below the model equilibrium
    curve, the simulated plasma cools down and the average potential energy
    increases. Oppositely, there is a transfer from potential to kinetic
    energy when the initial configuration lies above the equilibrium curve.
    A remarkable agreement between the equilibrium curve obtained from the
    analytical model and the one defined by simulations runs is observed. 

    Figure~\ref{modsim_alpha} shows the ionization degree as a function
    of temperature at equilibrium. Two different simulation groups are
    plotted, for $V_i=4.75{\cal E}_0$ and $V_i=6.80{\cal E}_0$,
    respectively. Overall, when comparing with model predictions ---i.e.
    Eq~(\ref{Keqmodel})---, a good agreement is observed, with simulated
    ionization degree slightly overestimating model results. Differences
    come from the way in which electrons are classified as bound or free
    in either the model or the numerical simulation. In the theoretical
    model, a chemical picture is inherently used, and plasma constituents
    are viewed as atoms, ions and free electrons. Hence, bound and free
    electrons are clearly distinguished. MD simulations,
    however, naturally develop within a physical picture, where
    interactions among particles are treated on an equal many-body
    footing. Frontier between bound and free electron concepts is not so
    well defined and a criterion, like the one described above, must be
    specified to perform such classification and determine the ionization
    degree. From a comparison like the one shown in
    Fig.~\ref{modsim_alpha}, a numerical criterion  to match with the
    results that arise from the chemical picture implicitly assumed in
    the model could be extracted. However, as discussed below, this would
    require a deeper investigation about existence and treatment of
    \textit{collectivized electrons} in the plasma, which is beyond the
    scope of this work. 

\begin{figure}[t]
  \begin{center}
    \includegraphics[width=\columnwidth]{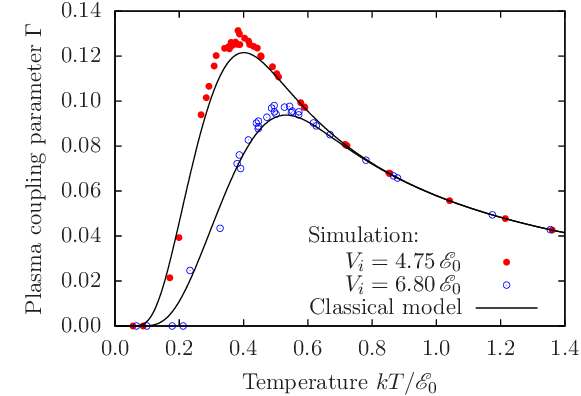}
    \caption{\label{modsim_gammaexp}
    	Plasma coupling parameter as a function of temperature at
	equilibrium. Comparisons between the analytical model and
	simulations are shown for two cases, i.e.
	$V_i=4.75{\cal E}_0$ and $V_i=6.80{\cal E}_0$ (with $\Gamma_E=0.116$).}
  \end{center}
\end{figure}

\begin{figure}[b]
  \begin{center}
    \includegraphics[width=\columnwidth]{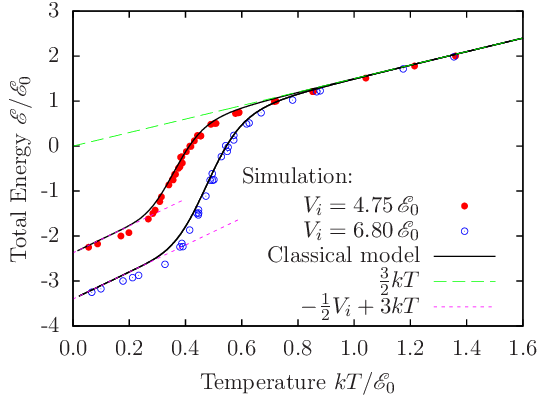}
    \caption{\label{modsim_etot}
    	Total energy per particle as a function of temperature at
	equilibrium. Comparisons between the analytical model and
	simulations are shown for two cases, i.e.
	$V_i=4.75{\cal E}_0$ and $V_i=6.80{\cal E}_0$ (with $\Gamma_E=0.116$).}
  \end{center}
\end{figure}

    Once the equilibrium is reached, corresponding temperature, ionization
    degree and free electron density values are obtained, and therefore the
    \emph{experimental} plasma coupling parameter ---i.e. as result of the
    \emph{computational experiment}--- can be determined.  Simulation results
    shown in Figs.~\ref{modsim_eqcurve} and \ref{modsim_alpha}, using a
    fixed numerical input parameter $\Gamma_E=0.116$, lead to
    \emph{experimental} coupling parameter values in the range
    $0\lesssim\Gamma_{exp}\lesssim0.13$ ---see Fig.~\ref{modsim_gammaexp}---,
    thus spanning from weakly- to moderately-coupled plasmas.
    We here recall the physical meaning and difference between dimensionless
    parameters $\Gamma_E$ and $\Gamma$, as described above in
    Sec.~\ref{pardyn}. Numerical parameter $\Gamma_E$ controls the strength
    of particle interactions ---see Eq.~\ref{elecmeq}, which is written in
    simulation units---. In other words, $\Gamma_E$ defines the magnitude
    of electron charge in the simulation. Coupling parameter $\Gamma$
    instead is determined by the ratio between the typical free electron
    interaction energy and the corresponding kinetic energy, i.e. it gives
    a relation between free electron density and electron temperature, and
    therefore accounts for the system \textit{coupling degree}. When a
    simulation is launched, a certain amount of energy is delivered to the
    system by means of draw-resulting initial conditions. The system
    consistently evolves according to interaction strength fixed by
    $\Gamma_E$ and energy is redistributed to eventually reached an
    equilibrium state and ionization balance with well-defined values for
    kinetic and potential energy. At such equilibrium state we can
    \textit{measure} the resulting free electron density and temperature,
    and therefore obtain $\Gamma_{exp}$.
    As observed in Fig.~\ref{modsim_gammaexp}, in the low 
    temperature limit, plasma mainly consists of non-interacting neutral
    pairs, and consequently coupling is weak. As temperature rises, plasma
    ionizes which favors the ion-electron coupling up to a maximum value.
    If temperature further increases, kinetic energy clearly overcomes the
    potential contribution, and coupling between particles drops. In the
    moderate-coupling range, interactions among particles in fact play a
    non negligible role
    and MD calculations are therefore meaningful. For the sake of
    completeness, we checked that our simulation technique is robust
    enough to deal with strong-coupling conditions, and a good agreement
    between simulation and model was also found for coupling parameter
    values up to $\Gamma_{exp}\approx1$. Above this limit classical picture
    becomes questionable since the characteristic
    atomic size may get comparable to the average electron distance.
    Also, as expected, results from MD simulations reproduce
    those from IPA simulations for the case of weakly coupled plasmas,
    with both comparing well with the statistical model.

    At equilibrium, total energy behavior as a function of temperature
    is easy to interpret ---see Fig.~\ref{modsim_etot}---. At low
    temperatures, system mainly consists of neutral atoms, which have a
    total of 6 degrees of freedom ---i.e. 3 translational degrees plus
    3 internal degrees---, and therefore total energy goes as $3kT$.
    In the high-temperature limit, plasma is fully ionized, it behaves
    like a monoatomic ideal gas and total energy per particle equals
    the translational kinetic energy, i.e. $\frac{3}{2}kT$.
    In the intermediate regime we observe the \textit{phase transition}
    between neutral and fully ionized plasma states. In this region,
    total energy per particle strongly increases with temperature. In
    other words, the required total energy to produce a temperature
    increasing is significant, because most of it is employed to overcome
    the neutral atom binding energy.  

\begin{figure}[t]
  \begin{center}
    \includegraphics[width=\columnwidth]{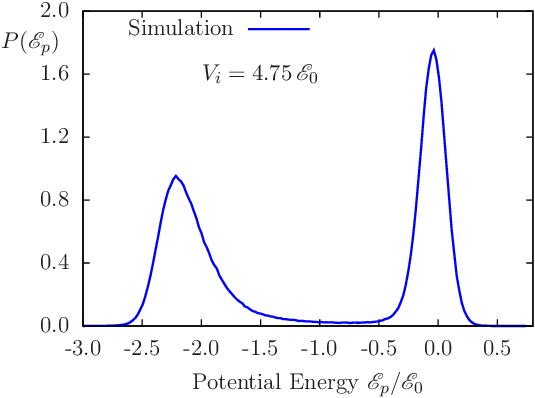}
    \caption{\label{potener_dist}
    	Statistical distribution of electron potential energy (per
	particle) obtained from a simulation at equilibrium. For the
	case shown, equilibrium conditions are $kT=0.37{\cal E}_0$
	and $\alpha=0.53$.}
  \end{center}
\end{figure}

    Further illustration of this \textit{phase equilibrium} is shown in
    Fig.~\ref{potener_dist}, where we plot the statistical distribution
    of potential energy per particle at particular equilibrium conditions.
    In the example, plasma ionization results $\sim 50\%$, so that
    potential energy distribution shows two well separated bell-shaped
    peaks having similar area values and arising from the bound ---left
    one--- and free ---right one--- electron ensembles. Moreover, it is
    seen that the average potential energy of \textit{free} electrons
    lies slightly below zero. This \textit{plasma collective effect}
    naturally emerges from the simulation and reflects the fact that many
    electrons in the plasma have a negative total energy and are not bound
    to a particular positive ion but to the plasma as a whole, hence they
    sometimes are referred as collectivized
    electrons~\cite{fisher2002,lankin2008,lankin2009}.
    This behavior has to do with the effect of \textit{ionization
    potential depression} (IPD)~\cite{fisher2002,crowley2014} that appears in
    dense plasmas and also illustrates how diverse the information provided
    by MD simulations can be. In fact, some IPD investigations in aluminum
    plasmas at and out of thermodynamic equilibrium have already been
    performed by using multi-component classical MD
    simulations~\cite{calisti2015,calisti2015b}.

    Overall, results shown in this section reveal robustness of numerical
    algorithms implemented in the code, confirm a proper description of
    ionization-recombination mechanism and ultimately provide confidence
    on the physics that can be extracted from our MD simulations. Once
    simulation technique has been validated, the analytical statistical
    model now becomes a useful tool to \textit{(a)} anticipate the
    simulation equilibrium state and thus securely interrupt a
    simulation run and \textit{(b)} optimize the calculation by
    consistently improving the choice of initial configuration when
    launching a simulation.  

\section{Conclusions}

    Particle dynamics simulations of hydrogen plasmas have been performed
    in the context of classical molecular dynamics. Theoretical basis of
    simulation model as well as numerically relevant aspects are
    discussed in detail, thus proving a thorough implementation of the
    computer simulation technique. Particle dynamics equations are solved
    without using any thermostat algorithm and the simulation model properly
    deals with the ionization-recombination mechanism. A comprehensive study
    of equilibration process is made, with emphasis
    on the need of reaching the stationary stage for a safe statistical
    sampling of relevant physical quantities. Molecular dynamics
    simulations are often considered as idealized experiments, where
    different effects can be artificially switch on and off to assess
    their potential impact, thus providing a deep insight of the
    underlying physics and a unique testbed for theory validation.
    However, these simulations are certainly challenging and consequently a
    validation process is also demanded. Here we developed an analytical
    statistical equilibrium model for computational validity
    assessment of plasma particle dynamics simulations. A good
    agreement between model and molecular dynamics results was obtained
    in a wide range of plasma coupling parameter, thereby revealing the
    robustness of employed numerical algorithms and ultimately providing
    confidence on the physics that can be inferred from simulation
    results. Continuing with the research on plasma Stark-broadening performed
    by our group over the last three decades, the internal consistency and
    validity tests of MD simulations performed in this work are a first step
    for the ultimate goal of carrying out a detailed investigation of the
    impact of ionization-recombination dynamics on  broadening mechanisms
    of spectral line shapes from emitting ions in multi-component plasmas 
    ---i.e. beyond a fully ionized scenario---. This topic will be addressed
    in a forthcoming publication.

\appendix*

\section{Some considerations on numerical heating}
\label{AP1}

    It is well known that combined and accumulated effect of computational
    errors in molecular dynamics leads to the so-called
    \textit{numerical heating}~\citep{hockneybook,ueda1994,rambo1997,evstatiev2013}.
    In fact, the effect manifests as a total energy increase since, as it can
    easily demonstrated, such \textit{heating} not always results in a temperature
    rise. Here, we will discuss this phenomenon in order to assess both potential
    imposed limitations on molecular dynamics and its practical consequences when
    using simulation techniques for the statistical analysis of local electric
    microfield. 

    At the time instant $t$, total energy of a system configuration is given by  
    \begin{equation}\label{EQ35}
	{\cal E} = \sum_{i}\,\frac{1}{2}m_iv_i^2
		+\frac{1}{2}\sum_{{i,j}\atop{i\neq j}} V(r_{ij}),
    \end{equation}
    with $\vecr_{ij} = \vecr_i-\vecr_j$. After each time step $\Delta t$,
    $\vecv_i$ and $\vecr_i$ will change, thus producing the consequent energy
    variation. Taking into account that, numerically,
    $\vecv(t+\Delta t) = \vecv(t)+\frac{1}{m}\vF\,\Delta t$, then we have
    \begin{eqnarray}
    	\label{EQ36}
	    \Delta v_i^2 & = & v_i^2(t+\Delta t) - v_i^2(t) \nonumber
	\\ & = & 2\frac{1}{m_i}\vecv_i\cdot\vF_i\Delta t
		+\frac{1}{m_i^2}F_i^2\Delta t^2,		
	\\ \label{EQ37}
	    \Delta V(r_{ij}) & \approx & \left(
	    	\frac{\partial V}{\partial\vecr_{ij}}
		\right) \cdot\Delta\vecr_{ij} \nonumber
	\\ & = & -\vF_{ij}\cdot\vecv_{ij}\Delta t \nonumber
	\\ & = & -\vF_{ij}\cdot(\vecv_i-\vecv_j)\Delta t,
	\nonumber
    \end{eqnarray}
    so it follows
    \begin{eqnarray}\label{EQ38}
	\Delta{\cal E} & \approx & \sum_{i}\,
		\vecv_i\cdot\vF_i\,\Delta t
		+\frac{1}{2}\sum_{i}\,\frac{1}{m_i}F_i^2\Delta t^2\;
		-\;\nonumber
	\\ & & - \frac{1}{2}\sum_{{i,j}\atop{i\neq j}}
		\vF_{ij}\cdot(\vecv_i-\vecv_j)\Delta t.
    \end{eqnarray}
    Also,
    \begin{equation}\label{EQ39}
	\vF_i = \sum_{j\atop{j\neq i}} \vF_{ij},
		\hspace{2em}{\rm with}\hspace{1em}
		\vF_{ij} = -\vF_{ji}.
    \end{equation}
    Then, first and third addends in Eq.~\Ref{EQ38} cancel out and it is found
    \begin{equation}\label{EQ40}
	\Delta{\cal E} \approx \sum_{i}\,
		\frac{1}{m_i}F_i^2\Delta t^2,
    \end{equation}
    which is a positive quantity.  
    
    Therefore, it is said that system numerically heats up, but it does not
    necessarily mean that kinetic and potential energy separately increase.
    It will depend on the system configuration, i.e. depending on particles
    \textit{location}, either kinetic or potential energy could even decrease,
    but their sum will always increase.

\begin{figure}[h]
  \begin{center}
    \includegraphics[width=\columnwidth]{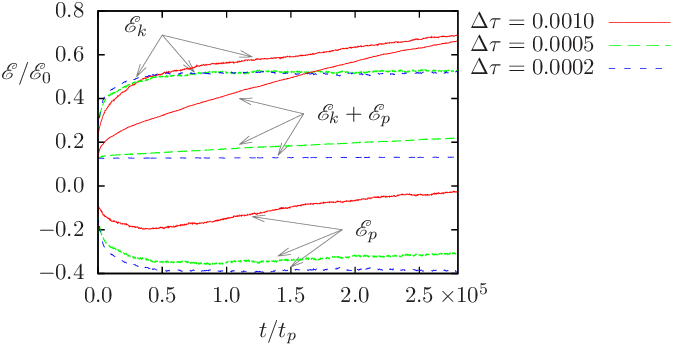}
    \caption{\label{calen1}
       Time histories of kinetic, potential and total energy for three
       simulation cases. In order to assess the impact of numerical heating,
       different time step values, $\Delta t$, were used in each case to
       solve the system of motion equations.}
  \end{center}
\end{figure}

    We illustrate this fact in Fig.~\ref{calen1}. Time evolution of kinetic,
    potential and total energy are shown for three simulation cases ---hereafter,
    it should be noted that each case actually represents an average over 8
    independent simulation runs--- that were executed using different time step
    values. The three cases started from exactly the same initial configuration.
    When looking at the figure it might seem simulations start at different
    points, but this is because the typical sudden initial change happens in a
    different way for each case ---note that simulation early times have not
    been zoomed in the figure---. For the greatest time step value numerical
    heating is certainly noticeable, whereas for the smallest one is negligible.
    Also, under certain conditions, it is observed that numerical heating leads
    to an increase of potential energy per particle, while temperature remains
    constant ---see the case for $\Delta t = 5\times10^{-4}t_p$---.  

\begin{figure}[h]
  \begin{center}
    \includegraphics[width=\columnwidth]{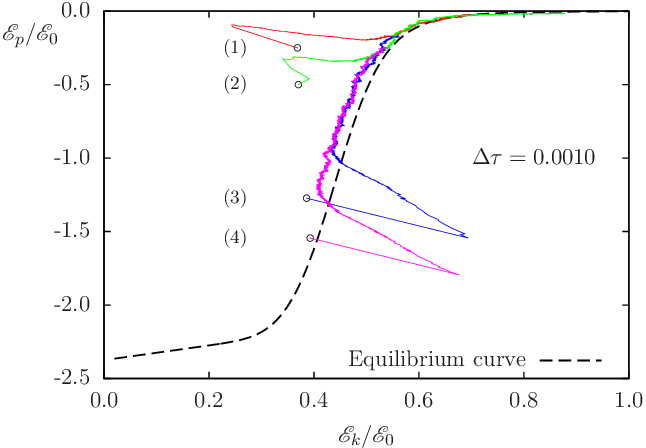}
    \caption{\label{calen2}
	Simulation trajectories in the ${\cal E}_p$:${\cal E}_k$ plane.
	Four simulation cases launched from different initial configuration
	are shown. Pointedly, a coarse time step was chosen, so that numerical
	heating can not be avoided and total energy increases with time in each
	case.} 
  \end{center}
\end{figure}

    Impact of numerical heating can be deeper analyze by means of
    Fig.~\ref{calen2}, where trajectories in the ${\cal E}_p$:${\cal E}_k$
    plane are shown for four simulation cases launched from different initial
    conditions. In all of them the same relatively coarse time step was employed.
    As already mentioned, early in time a sudden exchange between kinetic and
    potential energy takes place. Then, the system faces its approach to
    equilibrium with the total energy continuously increasing, i.e. unlike
    observed in Fig.\ref{modsim_eqcurve} and as a consequence of numerical
    heating, for the cases shown in Fig.~\ref{calen2} path to equilibrium does
    not follow a ${\cal E}_k+{\cal E}_p=cte$ trajectory.  In labeled cases (3)
    and (4), relaxation phase develops as plasma cools down ---i.e. kinetic
    energy per particle decreases---,  so that the increase in total energy
    due to numerical heating actually manifests as a potential energy increase.  

    Once the system reaches the equilibrium curve ---the one obtained from the
    analytical model discussed in Sec.~\ref{eqmodel} is displayed for
    reference---, such total energy increase distributes among kinetic and
    potential energy. Still, it mainly entails an increase of potential energy,
    since temperature does not change too much in the process ---note that
    different scales are used in kinetic and potential energy axes---. Thus,
    if a thermostat algorithm ---i.e. a temperature control--- were used to
    force the simulation to stabilize, one might not realize about numerical
    heating because the numerically-added energy amount would mostly turn into
    potential energy, with the kinetic one barely changing in the simulation.
    Nevertheless, strictly speaking, when the system begins to move following
    the equilibrium curve numerical heating truly leads to a temperature
    increase. In fact, if the system is left to evolve for a long time it
    will end up in the fully-ionized state. We note in passing that this
    picture is not consistent with a steady ionization-recombination equilibrium
    state.

    On the other hand, since after certain time the system evolves going
    through subsequent equilibrium configurations, if numerical heating slowly
    builds up its effect will be tolerable provided that desired information
    from the simulation does not require the extraction of long time-histories.

    Nevertheless, it should be kept in mind that heating effect may have a
    greater impact on particle spatial distribution than on plasma temperature
    itself, so a separate surveillance of kinetic and potential energy is needed
    as well as a consequent handling of simulation data. In this connection,
    we recall that average potential energy arises from particle spatial
    arrangement, which ultimately determines the microfield statistical
    distribution. As known, the latter plays a pivotal role in the study
    of Stark effect. 

    Analysis shown here guided the calculations performed for this work.
    Thus, as mentioned, in all simulation runs discussed in
    Sec.~\ref{results}, time step was chosen small enough to make negligible
    the numerical heating effect.

\begin{acknowledgments}
    This work has been supported by Research Grants ENE2013-42581-R/FTN
    and ENE2015-67581-R/FTN (MINECO/FEDER - UE) from the Spanish
    Ministry of Economy and Competitiveness. This work has been
    carried out within the framework of the EUROfusion Consortium and
    has received funding from the Euratom research and training
    programs 2014-2018 under grant agreement No 633053 (Projects
    AWP17-ENR-IFE-CEA-01 and AWP17-ENR-IFE-CEA-02). The views and
    opinions expressed herein do not necessarily reflect those of
    the European Commission.
\end{acknowledgments}


\bibliography{bibliography.bib}

\end{document}